\documentstyle[12pt]{article}

\newcommand{\newsection}[1]{
\vspace{15mm} \pagebreak[3] \addtocounter{section}{1}
\setcounter{subsection}{0} \setcounter{footnote}{0}
\noindent {\large\bf \thesection. #1} \nopagebreak
\medskip
\nopagebreak}

\newlength{\extraspace}
\newlength{\extraspaces}
\newcommand{\ba}{\begin{eqnarray}
\addtolength{\abovedisplayskip}{\extraspaces}
\addtolength{\belowdisplayskip}{\extraspaces}
\addtolength{\abovedisplayshortskip}{\extraspace}
\addtolength{\belowdisplayshortskip}{\extraspace}}
\newcommand{\ea}{\end{eqnarray}}
\newcommand{\be}{\begin{equation}
\addtolength{\abovedisplayskip}{\extraspaces}
\addtolength{\belowdisplayskip}{\extraspaces}
\addtolength{\abovedisplayshortskip}{\extraspace}
\addtolength{\belowdisplayshortskip}{\extraspace}}
\newcommand{\ee}{\end{equation}}
\newcommand\com[2]{{\bigl[{#1},{#2}\bigr]}}
\newcommand\anti[2]{{\bigl\{{#1},{#2}\bigr\}}}
\newcommand{\zbar}{{\bar z}}
\newcommand{\Xbar}{\mbox{\small $\bar Z$}}
\newcommand{\XX}{\mbox{\small $Z$}}
\newcommand{\TTheta}{\mbox{\small $\Theta$}}
\newcommand{\thetabar}{{\bar \theta}}
\newcommand{\Thetabar}{{{\mbox {\small $\bar \Theta$}}}}
\newcommand{\PP}{{\mbox{\small \bf H}}}
\newcommand{\QQ}{{\mbox{\small \bf Q}}}
\newcommand{\SS}{{\mbox{\small \bf S}}}
\newcommand{\HH}{{\mbox{\small \bf D}}}
\newcommand{\KK}{{\mbox{\small \bf K}}}
\newcommand{\JJ}{{\mbox{\small \bf J}}}

\newcommand{\Tr}{{\rm Tr}}

\def\bea{\begin{eqnarray}}
\def\eea{\end{eqnarray}}
\def\appendix#1{
  \addtocounter{section}{1}
  \setcounter{equation}{0}
  \renewcommand{\thesection}{\Alph{section}}
  \section*{Appendix \thesection\protect\indent \parbox[t]{11.15cm}
  {#1} }
  \addcontentsline{toc}{section}{Appendix \thesection\ \ \ #1}
  }
\newcommand{\half}{{1\over 2}}
\newcommand{\CH}{{\cal H}}

\newcommand{\CA}{{\cal A}}

\def \is {\! & \! = \! & \! }

\def \ppi {{\mbox {\large $\pi$}}}
\def \pppi {{\mbox {\large $\pi$}}}

\begin{document}

\addtolength{\baselineskip}{.2mm}

\begin{titlepage}
\begin{center}

{\hbox to\hsize{ \hfill }} {\hbox to\hsize{ \hfill PUPT/2113}}

{\hbox to\hsize{ \hfill hep-th/0403024}}

\vspace{2.5cm}

{\Large \sc 
Superstrings on $AdS_2$
}\\[4mm]{\Large \sc
and}\\[4mm]
{\Large \sc Superconformal Matrix Quantum Mechanics}\\[1.5cm]

{\large Herman Verlinde}\\[8mm]

{\large \it Department of Physics\\[2mm]
 Princeton University\\[2mm]
Princeton, NJ 08544}\\[6mm]

\vspace*{18mm}

{\large\sc Abstract}\\

\end{center}
\noindent We construct a $\kappa$-symmetric Green-Schwarz action
for type IIA string theory on $AdS_2$.
As a candidate holographic dual, we
consider superconformal matrix quantum mechanics, given by
the Marinari-Parisi model with vanishing or logarithmic
superpotential. We derive that the super-eigenvalues form
a consistent subsector, and that their dynamics
reduces to that of the supersymmetric Calogero-Moser model. The
classical string action and the matrix model both have an
infinite set of conserved charges, that include an ${\cal N}\!=\!2$
target space
super-Virasoro symmetry. 
As a microscopic test of the duality, we reproduce the exact form
of the Calogero interaction via a string worldsheet calculation.


\end{titlepage}

\newsection{Introduction}

The study of low-dimensional non-critical string theories has produced
various useful lessons about string theory in higher dimensions.
In particular, their matrix model formulations provide explicit
examples of the holographic duality between non-abelian dynamics
on D-branes and closed string theory in one higher
dimension.  Most of the recent work on non-cricital string theory
thus far concentrates on bosonic strings or type 0A or 0B superstrings,
since these are best understood
\cite{reloaded}\cite{recent}\cite{review}\cite{typezero}\cite{hat}.

Much less is known about non-critical IIA or IIB superstrings \cite{davnat}.
D-branes in this theory are only beginning to be explored \cite{tohru}\cite{ahn},
and a matrix formulation still needs to be uncovered\footnote{For
a (still incomplete) proposal see \cite{mp}\cite{atish}\cite{super}.}.
There are several reasons why progress on this front has been slow.
First, the NS-R worldsheet CFT arises from ${\cal N}\!=\! 2$ Liouville
theory via a chiral GSO projection, that strictly
exists only in a euclidean target space \cite{davnat}.
Secondly, because target-space supersymmetry is broken by the
linear dilaton background, it is a non-trivial task to construct
a Green-Schwarz world-sheet action \cite{gs} and extract the full
supersymmetric
form of the target-space effective action. Finally, general type II
backgrounds carry RR flux, and only limited tools are available for
studying strings in such backgrounds.
Nonetheless, in spite of its underdeveloped state, it
is expected that 2-D type II superstring theory exists, and will
reveal interesting new structures.


The bosonic sector of 2D type IIA string theory has, in addition
to a metric and dilaton, a non-dynamical RR two-form flux $F=dA$.
Their low-energy equations of motion are summarized by the
effective action \be \label{seff} S_2 = \int \! d^2 x\,
\sqrt{-g}\Bigl(e^{-2\phi}(R + 4(\nabla \phi)^2 + 8) - F^2\Bigr)
\ee This system was studied in
\cite{bo}\cite{Polyakov2000}\cite{gtt}, where it was shown to
admit extremal black hole solutions with non-zero flux, and with a
near-horizon $AdS_2$ region. Concretely, the near-horizon solution
takes the form \be \label{adst} e^{2\phi} = {1\over q^2}\, ,
\qquad A_\pm = {q\over 2(z^+-z^-)}\, , \qquad ds^2_2 = - {dz^+
dz^- \over 4 (z^+-z^-)^2}\, . \ee Here $q$ parametrizes the
quantized RR-flux. The $AdS_2$ is a natural region to focus on,
because it is the most symmetric target space for the non-critical
string. In particular, target-space supersymmetry is likely to be
unbroken. However, while (\ref{adst}) indeed solves the equations
of motion of (\ref{seff}), the solution is not really valid within
its regime of validity: the curvature of the AdS-space is $R=-8$
in string units, and $\alpha'$-corrections to (\ref{seff}) are
non-negligible. A proper CFT description of this background would
be needed to establish that the solution actually persists after
the string corrections are taken into account. We will address
this problem in the section 2.

Assuming that IIA superstring theory in $AdS_2$ exists,  it is
reasonable to expect that the theory admits a holographic dual in
terms of a suitable quantum mechanical matrix model, with an
invariance  under a supersymmetric extension of the $AdS_2$
isometry group \cite{Gibbons}\cite{Britto}\cite{andyl}. Moreover, it seems
likely that the dual theory has a microscopic interpretation as
the world-line theory of a bound state of many D-particles of the
IIA theory. This suggest that the relation between IIA on $AdS_2$
and its dual represents an interesting triple intersection
point of AdS/CFT, old matrix models, and M(atrix) theory.

This paper is organized as follows. In section 2 we construct a
$\kappa$-symmetric Green-Schwarz action for $AdS_2$ in terms of an
$Osp(2|1)/SO(2)$ coset model. The world-sheet action is an exactly
soluble field theory. By studying the region near the $AdS_2$
boundary, we find evidence that the model possesses a target space
${\cal N}\!=\!2$ super-Virasoro symmetry. In section 4, following
ideas of Strominger, Gibbons and Townsend,
\cite{Gibbons}\cite{andyl} we formulate a candidate holographic
dual of the IIA theory on $AdS_2$, in the form of superconformal
matrix quantum mechanics: a Marinari-Parisi matrix model with a
superpotential $W(\Phi)= q\log \Phi$. We show that the matrix
eigenvalues form a consistent supersymmetric sub-sector and are
described by the supersymmetric Calogero-Moser model, which (for
$q=0$) also exhibits an ${\cal N}=2$ super-Virasoro symmetry
\cite{bergvas}. Finally, in section 6, we outline a possible
microscopic string derivation of the Calogero interaction in terms
of a Knizhnik-Zamolodchikov equation of a worldsheet $gl(1|1)$
current algebra.

\newsection{GS String Action on $AdS_2$}

\noindent We will regard the field space of the IIA theory
on $AdS_2$ as a coset manifold ${Osp(1 | 2)/SO(2)}$. The global
symmetry group $Osp(1|2)$ is a
supergroup with 3 bosonic generators
$(\PP,\HH,\KK)$ and 2 fermionic generators
$(\QQ^{},\SS^{})$. The commutators are \cite{Britto}
\ba \com{\PP}{\KK} = \HH,
\qquad \quad & \com{\PP}{\HH} = -2\PP, & \qquad \com{\KK}{\HH} =
2\KK,
\nonumber \\[1mm]
\anti{\QQ^{}}{\QQ} = -2 \, \PP, \; \qquad & \com{\QQ^{}}{\HH} =
-\QQ^{},& \qquad  \com{\QQ^{}}{\KK} = \SS^{},
\\[1mm]
\anti{\SS^{}}{\SS} = 2\, \KK,\ \qquad  & \com{\SS^{}}{\HH} =
\SS^{},& \qquad  \com{\SS^{}}{\PP}
= \QQ^{}, \nonumber \\[1mm]
\anti{\SS^{}}{\QQ} = - \, \HH,\, \qquad & \com{\QQ^{}}{\PP} = 0,&
\qquad \com{\SS^{}}{\KK} = 0.
\nonumber \ea 
The $SO(2)$ factor that is divided out is generated by the
dilation operator $\HH$.

\bigskip
\bigskip

\noindent
{\sc GS action in terms of 1-forms}

The most convenient notation for obtaining the supersymmetric and
$\kappa$-invariant string action is to use the formalism of Cartan
1-forms \cite{henneaux}\cite{metsaev}.
We will denote the left-invariant Cartan 1-forms by the same
letters as the corresponding generators, that is, $H$ denotes the
$\PP$-component of $G^{-1} dG$, etc. The resulting Maurer-Cartan
equations are
\ba
\label{mc}
d H&\! =\!  & \ - 2 D \wedge {H} + \,  
Q^{} \wedge Q_{}\nonumber\\[1mm]
d D & \! =\! & \ \ {H} \wedge K \; + \;
Q^{}
\wedge S_{} \nonumber\\[1mm]
d K & \! = \! & \;  2 D \wedge K \, -\, 
\, S^{}  \wedge S_{} \\[1mm]
d Q^{} & \! =\! & \ -  D \wedge Q^{} + \; {H} \wedge S^{}
\nonumber\\[1mm]
d S^{} & \! =\! & \; D\wedge S^{}  + \, K \wedge Q^{}
.
\nonumber \ea
One way to read these equations is that, when viewed as 1-forms on
the worldsheet, the Cartan forms combine into a flat
$Osp(1|2)$-connection.

The $\kappa$-symmetric string
action is the sum of a kinetic and Wess-Zumino term
\cite{henneaux}\cite{metsaev}:
\be {\cal S}= - {1\over 2}
\int\limits_{{\cal M}_2 } \! d^2\sigma \sqrt{g} \, g^{\alpha\beta}\! H_\alpha K_\beta
\; + \; WZ \,. \ee
The WZ term is almost uniquely determined by invariance under the
$SO(2)$ subgroup. It can be written as (here $\partial {\cal M}_3
= {\cal
M}_2$) \be WZ = 
\int\limits_{{\cal M}_3} ( Q^{} \wedge Q_{} \wedge K -  S^{}
\wedge S_{}  \wedge H) \,= \frac{1}{2} \int\limits_{{\cal M}_3}
d(H\wedge K), \ee where we used (\ref{mc}). The coefficient of the
WZ term is fixed by $\kappa$-symmetry.

The string action thus simply reduces to \be \label{gsform} {\cal
S} = -{1\over 2}\int\limits_{{\cal M}_2 } \! d^2\sigma (\sqrt{g}
\, g^{\alpha\beta}\! + \epsilon^{\alpha\beta} ) H_\alpha K_\beta =
- \int\limits_{{\cal M}_2 } \! d^2z \, H_{\zbar} K_z\ee where $z$
and $\zbar$ are conformal gauge coordinates. Before verifying the
$\kappa$-symmetry of this action, we will first write it in a more
familiar form.

\bigskip
\bigskip

\noindent
{\sc GS action as a sigma-model}


Let us introduce a
concrete parameterization of the Cartan 1-forms in terms of target
superspace coordinates $(\XX,\Xbar,\TTheta^{},\Thetabar^{})$ as
follows:
{\renewcommand{\small}{\footnotesize}
\ba \label{param} H= 
\frac{d\XX_{}+\TTheta^{} d\TTheta} {\XX_{}-\Xbar_{}- \TTheta^{}
\Thetabar^{}}
\, , \ \ \ \qquad & &  \qquad  \ \ K  = 
\frac{d\Xbar_{}+\Thetabar^{} d\Thetabar_{}}
{\XX_{}-\Xbar_{}- \TTheta_{}^{}\Thetabar^{}} \nonumber \\[2.5mm]
Q =
\frac{d \TTheta-  ( \TTheta_{}^{}\!\! -\Thetabar) \, H}
{\sqrt{\XX-\Xbar - \TTheta^{}\Thetabar_{}^{}}} \, , \qquad & & \qquad
S^{} =
\frac{d \Thetabar^{} +  ( \TTheta^{}\!\! -\Thetabar) \, K}
{\sqrt{\XX-\Xbar - \TTheta^{} \Thetabar_{}^{}}} \ea \ba D =
\frac{i(d\XX_{}+d\Xbar_{}+ \Thetabar^{} d\TTheta+\TTheta^{}
d\Thetabar)} {\XX_{}-\Xbar_{}- \TTheta^{}\Thetabar}. \nonumber
\ea}
These formulas represent a flat  $Osp(1|2)$ connection
$A=-g^{-1}dg$ with \cite{ezawa}
{ \small \renewcommand{\small}{\footnotesize}
\be g= {1\over \Xi} \left(\begin{array}{ccc}
\XX_{} \sqrt{\Xi} &
\Xbar_{} \sqrt{\Xi} &
(\Thetabar_{}\XX\! -\TTheta_{}\Xbar_{})\\
\sqrt{\Xi} &
\sqrt{\Xi} &
(\Thetabar_{}-\TTheta_{})\\
\TTheta_{}\sqrt{\Xi} &
\Thetabar_{} \sqrt{\Xi}& \Xi
-\TTheta_{}\Thetabar_{}
\end{array}\right)\, ,
\qquad \quad \Xi \equiv \XX_{}-\Xbar_{}- \TTheta_{}\Thetabar_{}\, .
\ee}
All fields in this espression are
worldsheet scalars with spin (0,0).

Inserting (\ref{param}) in (\ref{gsform}),
the GS action reduces to a more familiar looking sigma-model,
with the Poincare super-upperhalf plane as target space
\be \label{gssig} {\cal S} = - \int\limits^{}_{}\! d^2 z \,
\frac{(\bar\partial\XX \!+\TTheta^{} \bar\partial \TTheta_{}
)(\partial \Xbar \! + \Thetabar^{}
\partial \Thetabar_{})}{(\, \XX - \Xbar\,  -\, \TTheta^{}
\Thetabar_{} \, )^2} \, .  \ee This action is, as it should,
invariant under the global symmetry group $Osp(1|2)$, which acts
via the super-M\"obius transformations {\renewcommand{\small}
{\footnotesize}\ba
\label{mobius}
\XX_{}&\rightarrow&\frac{a\XX_{}+b}{c\XX_{}+d}+ \frac{(-\delta^{}
\XX_{}+\epsilon^{})\TTheta_{}}{(c\XX_{}+d)^{2}}, \qquad \
\mbox{\small \ and \ \ C.C.}\nonumber
\\[2mm]
\TTheta^{}_{} & \rightarrow & \frac{\TTheta^{}}{c\XX_{}+d} +
\frac{-\delta^{} \XX_{}+\epsilon^{}}{c\XX_{}+d}\, , \qquad \qquad
\mbox{\small and \ \ C.C.} \ea} where $a,b,c,d\in{\bf R}$ are
Grassmann even constants and $\delta^{}, \epsilon^{}$ are
Grassmann odd constants, satisfying the relation
$ad-bc=1-\delta^{}\epsilon$.

Note that in (\ref{gssig}) we can recognize the presence of a
non-zero RR two-form: if we expand the
denominator in (\ref{gssig}) in $\Thetabar\TTheta$, we find a term
in the action proportional to the expected RR coupling \be
{\Thetabar\TTheta \bar\partial \XX \partial \Xbar  \over (\XX -
\Xbar)^3}\, . \ee

\bigskip

\bigskip

\noindent
{\sc Invariance under $\kappa$-symmetry}

We will now verify, following \cite{metsaev}, the invariance under local
$\kappa$-symmetry transformations. These act both on the target
space coordinates $X^M = (\XX,\Xbar,\TTheta^{},\Thetabar^{})$ as
well as on the worldsheet metric $g_{\alpha\beta}$. It is useful
to parameterize the $\kappa$-variations of the target space fields
in terms of the combinations \ba \delta_\kappa x \equiv H_{\!M}\,
\delta_\kappa X^M  \, , \qquad
 \quad \delta_\kappa {\bar x} \! &\! \equiv\! & \! K_M \, \delta_\kappa X^M
\qquad \quad \ \delta_\kappa x = \delta_\kappa {\bar x} = 0
\\[1.5mm]
\delta_\kappa \theta^{} \equiv Q^{}_M\, \delta_\kappa X^M \, ,
\qquad \quad \delta_\kappa \thetabar^{}  \! &\! \equiv\! & \!
S^{}_M \, \delta_\kappa X^M \, , \qquad \quad \delta_\kappa \alpha
\equiv D_{\! M}\,  \delta_\kappa X^M \nonumber \ea The condition
that $\delta_\kappa x = \delta_\kappa {\bar x} \!=\! 0$ translates
in terms of the sigma-model variables as
\ba \label{suco} \delta_\kappa \XX +
\TTheta^{} \delta_\kappa \TTheta_{} = 0 \, , \qquad \qquad
\delta_\kappa \Xbar + \Thetabar^{} \delta_\kappa \Thetabar_{} = 0
\, . \ea
Note that  this the condition for infinitesimal target space super-conformal
transformations. This observation will be of importance in the next section.

Using (\ref{suco}) and the explicit parameterization
(\ref{param}) of the Cartan 1-forms, a straightforward calculation
shows that $H$ and $K$ transform under $\kappa$-symmetry as \ba
\delta_\kappa H = {2 Q^{} } \, \delta_\kappa \theta_{}  +  H\,
\delta_\kappa \alpha \, \qquad \qquad \delta_\kappa K = {2 S^{} }
\, \delta_\kappa \thetabar_{}
 - K  \delta_\kappa \alpha
\ea It is now easy to verify that the action (\ref{gsform}) is
invariant under the local $\kappa$-variations  \cite{metsaev}
\ba \delta_\kappa
\theta_{} = 2 H_z \, \kappa_{}^z \qquad && \quad
\delta_\kappa(\sqrt{g} \, g^{zz}) = \, 8\,\sqrt{g} \, g^{z\zbar}
Q^{}_\zbar \kappa_{}^z \,
\, ,\nonumber \\[2mm]
\delta_\kappa \thetabar_{} = 2 K_{\bar z}\, \kappa_{}^\zbar \qquad
& &  \quad \delta_\kappa(\sqrt{g}\, g^{\zbar \zbar}) = \, 8\,
\sqrt{g}\, g^{z\zbar} S^{}_z \kappa_{}^\zbar\, , \ea which
comprise 4 local chiral fermionic symmetries. Hence
$\kappa$-invariance allows one to eliminate all local on-shell
fermionic degrees of freedom of the sigma-model. Conformal
symmetry eliminates all local on-shell bosonic degrees of freedom.
Superstrings on $AdS_2$ do not have any transverse oscillations.

\bigskip



\newsection{Target Space Symmetries}

The existence of a GS action, that is $\kappa$-symmetric and
classicically conformally invariant, is a strong indication that
IIA string theory on $AdS_2$ exists. The main task would be to
quantize the theory, and verify that it is a conformal field
theory with the critical value for the central charge. This last
requirement should fix the size of the $AdS$-curvature.

In this section we comment on some of the target-space symmetry
properties of the $AdS_2$ GS action, which we expect to be
helpful in constructing the quantum theory. Our discussion
here will be brief and schematic; we hope to present a
more detailed treatment in a future
paper \cite{sasha}.

\bigskip

\bigskip

\noindent
{\sc Conserved charges and integrability}

The GS string action, in either form (\ref{gsform}) or
(\ref{gssig}), fits within the category of (super) coset models
that are known to possess an infinite set of hidden conserved
charges \cite{schwarz}. After setting the anti-commuting fields to zero,
(\ref{gssig}) reduces to the $SL(2,R)/U(1)$ coset model
that features in the 2-dimensional reduction of 4-d gravity, which
is perhaps the oldest and most well-known classically integrable
field theory of this type.

In an interesting recent paper \cite{bena}, Bena, Polchinski and Roiban have
shown how to generalize the standard construction of the conserved
charges to $\kappa$-symmetric string actions. Our model indeed
exactly fits the analysis in section 3 of their paper.  In our
case, the consequences of this integrable structure for the
physical IIA string spectrum are somewhat limited by
the fact that conformal and $\kappa$-symmetry eliminate all local
world-sheet fluctuations. Still, the integrability will undoubtedly
be helpful in constructing the ground ring and for studying D-branes.

\bigskip
\bigskip

\noindent {\sc Free fields near AdS-boundary}

As a possibly helpful step in constructing the quantum theory, it
is convenient to add a one-form field $(\ppi,\bar \ppi)$ and rewrite
the kinetic term for $\XX$ and $\Xbar$ in first order form. One
obtains the following action \be \label{newsig} {\cal S} = \int \!
d^2 z \, \Bigl(\ppi \bar\partial \XX + \bar\ppi
\partial \Xbar  + \TTheta^+ \bar \partial \TTheta +
\Thetabar^+ \partial \Thetabar - \ppi\bar\ppi(\XX-\Xbar +
\Thetabar\TTheta)^2\Bigr) \ee where $\TTheta^+$ and $\Thetabar^+$
are defined as \be \label{cosnt} \TTheta^+ = \ppi \TTheta\, ,
\qquad \quad \Thetabar^+ = \ppi \Thetabar\, . \ee Eliminating
$\ppi$ and $\bar \ppi$ via their equation of motion, one recovers
(\ref{gssig}). At the quantum level, this step may in fact involve
some non-trivial renormalization, as well as the generation of a
linear dilaton term. The form of these quantum modifications is
constrained by $Osp(1|2)$-invariance.

The last term in (\ref{newsig}) becomes small near the boundary of
$AdS_2$. For the study of correlation functions that are dominated
by this boundary region, it would seem  reasonable to treat this
term as a perturbation. Setting it to zero leaves an apparently
free field theory, that, though still subject to the constraint
(\ref{cosnt}), decomposes into two chiral sectors. Each forms a free field
representation of an $Osp(1|2)$ current algebra (at level $k=-3/2$)
via \cite{wakimoto}:
\ba j^-(z) \is \ppi, \qquad \quad \ \qquad \qquad
\qquad \quad  s^-(z) =
\TTheta^+ + \ppi \TTheta, \nonumber \\[1.5mm]
j^3(z) \is \ppi\XX 
+ {1\over 2} \TTheta^+\TTheta \qquad \qquad \qquad
s^+(z) = -\XX(\TTheta^+ + \ppi \TTheta) 
+2 \partial \TTheta\, ,  \\[1.5mm]
j^+(z) \is \ppi \XX^2 
+ \XX \TTheta^+ \TTheta - {\textstyle{3\over 2}} \partial \XX +
{\textstyle{1\over 2}} \TTheta
\partial \TTheta \, .\nonumber
\ea The corresponding conserved global charges generate the
super-M\"obius transformations (\ref{mobius}). Although this free
field representation will certainly receive corrections from the
interactions, the appearance of a current algebra at this stage is
likely to be a helpful tool in constructing the full quantum
theory. Note that, already at the free field level, the presence
of the $AdS_2$ boundary implies that only the diagonal of the
$Osp(1|2)_L\times Osp(1|2)_R$ Kac-Moody symmetry survives.

\bigskip

\bigskip

\noindent
{\sc Super-Virasoro symmetry}

Let us explore the presence of this current algebra a bit
further. Given a set of $Osp(1|2)$ chiral currents, one can
construct an infinite set of conserved charges as follows \cite{gkn}
\ba L_n \is \oint \! dz \, [(n+1) j^3(z)
\XX^n(z) - n j^-(z) \XX^{n+1}(z)] \\[2mm] G_m \is \oint \! dz [
(m+\textstyle{1\over 2}) s^+(z)\XX^{m-\half}(z) -
(m-\textstyle{1\over 2}) s^-(z)\XX^{m+\half}(z)] \nonumber \ea
These charges all commute with the bosonic world-sheet conformal
algebra and are the analog of the DDF operators used in proving
the no-ghost theorem in critical string theory \cite{DDF}, and we
expect they can play the same useful role for our non-critical
model. As suggested by the notation, they generate a target-space
${\cal N}=1$ super-Virasoro algebra. Although our construction
above does not amount to an exact quantum treatment of the model,
we conjecture (based also on the integrability of the clsasical
theory) that this algebra survives at the full quantum level.

In light of our earlier observation below eqn (\ref{suco}), it is
reasonable to identify the $G_m$ with the $\kappa$-symmetry
generators. $\kappa$-Invariance thus implies that physical states
must be annihilated by the $G_m$ with $m\geq {1\over 2}$. In fact,
we can extend the
symmetry to a full ${\cal N}=2$ super-Virasoro algebra, by introducing
the charges \ba
\tilde{G}_{-{1\over 2}} \is \oint \! dz \, (\TTheta^+ - \ppi
\TTheta)\, , \qquad \quad \tilde{G}_{m-{1\over 2}} =
{\textstyle{2\over m+1}} \, [L_m,\tilde{G}_{-{1\over 2}}]
\nonumber\\[2mm] J_0 \is \oint \! dz \, \TTheta^+\Theta \qquad \qquad
\qquad \tilde{J}_{n} = {\textstyle{2\over n}} \, [L_n,J_0]
\ea The constraint (\ref{cosnt}) thus implies that $\tilde{G}_m$
vanishes on physical states. When combined, these observations are
evidence that the perturbative
spectrum of the string theory is obtained from an ${\cal N}=2$
superconformal target space field theory via the physical state
conditions \be \label{phys} G_m | {\rm phys}\rangle = 0\, , \qquad
\tilde{G}_m | {\rm phys}\rangle=0\, , \qquad L_m |{\rm
phys}\rangle = 0\, , \qquad J_m |{\rm phys}\rangle =0\, , \ee for
all $m>0$. The last two conditions are implied by the ${\cal N}=2$
algebra from the first two. Now for a moment it may
seem that one is left with very few target space degrees of
freedom. The states described by (\ref{phys}), however, are {\it
single} string states. We thus obtain an intriguing correspondence
between single string states and the primary states of the target
space ${\cal N}=2$ superconformal algebra.

\newcommand{\ppsi}{\hat J}

\newsection{Superconformal matrix quantum mechanics}

In this section we will formulate a candidate holographic dual of
superstring theory on $AdS_2$ in terms of superconformal matrix
quantum mechanics, which we propose has a microscopic
interpretation as the world-line theory of a bound state of many
D-particles of the non-critical IIA theory. Motivated by the
preceding discussion of the symmetries of the GS action, we will
consider a matrix model with manifest symmetry under the global
$Osp(2|2)$ subgroup of the ${\cal N}=2$ superconformal algebra. In
the next section we will show that this invariance can in fact
be extended to a full super-Virasoro algebra. The main technical
result of this section is a new derivation of the known result
\cite{atish}\cite{arod} that the eigenvalue dynamics of the supersymmetric
matrix model is described by a supersymmetric Calogero-Moser
model \cite{freedman}.

\bigskip

\bigskip

\noindent
{\sc Conformal Marinari-Parisi Model}

We introduce an $N\times N$ hermitian matrix superfield \be \Phi
\equiv \phi + \psi^\dagger \theta + \bar\theta\psi +
\bar\theta\theta F, \ee and write the supersymmetric action \be
\label{action} S = \int dt\,d\theta\,d\bar\theta\, {\rm Tr}\!
\left(\half \bar D \Phi D\Phi + W_0 (\Phi)\right). \ee This is the
action of the Marinari-Parisi model \cite{mp}. The quantum
mechanics of the matrix model is most conveniently described in a
Hamiltonian formalism. The conserved supercharges and hamiltonian
${\PP}= {1\over 2} \{ {\QQ}^\dagger,{\QQ}\}$ are \cite{arod}
{\small \ba { \bf Q}= \Tr\! \left\{\psi^\dagger\Bigl( \ppi + i
                {{\partial W_0(\phi)}\over{\partial \phi}}
\Bigr)\right\}\, , \qquad & & \qquad 
  {\bf Q}^\dagger = \Tr \!\left\{\psi\, \Bigl( \ppi - i
                {{\partial W_0( \phi)\over \partial \phi}}\Bigr)\right\},
\ea \ba \label{ham}
 {\bf H} \is \half\,{\rm Tr} \left(\ppi^2 + \Bigl({\partial W_0(
\phi)\over\partial
         \phi}\Bigr)^2\,
       \right)
       + \sum_{ijkl}\, [\psi^\dagger_{ij}, \psi_{kl}]\,
         {{\partial W_0( \phi)}\over{\partial \phi_{ij}\,\partial
\phi_{kl}}}, 
\ea} where $[\pppi_{ij}, \phi_{kl}] = -i\,\delta_{ik}\delta_{jl}$
and $\{\psi^\dagger_{ij},\psi_{kl}\} = \delta_{ik}\delta_{jl}$. We
will represent the Hilbert states as wavefunctions of $\phi$ and
$\psi$, so that \be \label{candu} \ppi_{ij} ={-i}{\partial \over
\partial \phi_{ij}} \qquad \qquad \psi_{ij}^\dagger = {\partial
\over \partial \psi_{ij}} \ee

In the original proposal of \cite{mp}, the superpotential
$W_0(\phi)$ was chosen to be a cubic function. If instead,
however, one considers a logarithmic potential
\be
W_0(\Phi) = q \log
\Phi,
\ee  the model has an additional scale invariance generated by
\be {\bf D} = {1\over 2}\Tr (\ppi \phi + \phi \ppi) \ . \ee
Combined with the supersymmetry algebra, this implies that the
matrix quantum mechanics is invariant under a full $Osp(2|2)$
symmetry, with as additional generators \ba {\bf S} = \Tr (\psi
\phi) \, , & \qquad &
{\bf S}^\dagger = \Tr(\psi^\dagger \phi)\, \nonumber\\[1.5mm]
{\bf K} = {1\over 2}\Tr (\phi^2) \, , & \qquad & {\bf J} = {1 \over 2}
\Tr(\psi^\dagger\psi)\, . \ea Given that we are after the
holographic dual of the IIA theory on $AdS_2$, this symmetry
algebra is a good place to start from. Although the general
super-conformal matrix model allows for a non-zero logarithmic
potential, we will for convenience restrict ourselves to the case
that $W_0=0$. Most of the following discussion is
straightforwardly generalized to the case with non-zero $W_0$.

In principle, by exploiting the superconformal supersymmetry, much
can be learned about the quantum mechanics of this model. To solve
for the general spectrum of excited states is still hard, due to
the interactions between the non-singlet degrees of freedom of the
matrix superfield. Via its proposed interpretation as the
world-line theory of IIA D-particles, however, we are motivated to
introduce an extra gauge invariance under conjugations, that will
enable us to eliminate some of these non-singlet degrees of
freedom. In fact, as we will now explain in some detail, it is
possible to introduce an auxiliary complex superfield $\CA$, such
that, via its equations of motion, the model is consistently
projected onto the supersymmetric singlet subspace. This fact that
the super-eigenvalues form a consistent supersymmetric subsector
of the MP matrix model was first observed by Dabholkar
\cite{atish}.

\bigskip

\bigskip

\noindent
{\sc From Marinari-Parisi to Calogero-Moser}

Using the auxiliary complex superfield $\CA$, we replace the
superderivatives in (\ref{action}) with gauge-covariant
superderivatives of the form \be \label{cova} D_{{}_{\!\! \CA}}
\Phi = D\Phi - [\CA,\Phi]\, , \qquad \quad \bar{D}_{{}_{\!\! \CA}}
\Phi = \bar{D}\Phi - [\bar{\CA},\Phi] \, . \ee These derivatives
are designed to be covariant under local gauge transformations
\be{ \Phi \mapsto e^{{\rm ad} \Lambda }~ \Phi \, , \qquad \quad
\CA \mapsto \CA + D\Lambda \, , \qquad \quad \bar{\CA} \mapsto
\bar{\CA} + \bar{D}\Lambda}\, , \ee with $\Lambda$ an arbitrary
real matrix superfield. Since $\CA$ and $\bar\CA$ appear as
non-dynamical fields, they can be eliminated via their respective
equations of motion
 \be
 \label{const}
[\Phi,\bar{D}_{{}_{\!\! \CA}}\Phi] = 0 \, , \qquad \qquad [\Phi,
{D}_{{}_{\!\! \CA}}\Phi]=0\, , \ee which shows that in the
classical theory, $\CA$ and $\bar \CA$ adjust themselves such that
the covariant derivatives (\ref{cova}) commute with $\Phi$. In
going to the quantum theory, the same equations become constraint
equations, which one may either want to impose before or after
quantization. We will choose the second route.

Let us introduce the canonical momentum superfields \be \Pi \equiv
{D}_{{}_{\!\! \CA}}\Phi\, , \qquad \qquad \bar\Pi \equiv {\bar
D}_{{}_{\!\! \CA}}\Phi. \ee If we choose our wave-functions to be
functions of $\phi$ and $\psi$, quantization gives that $\Pi$ and
$\bar \Pi$ act as \be \Pi = \psi + \theta \ppi 
\qquad \qquad \bar\Pi = \psi^\dagger + \bar \theta \ppi
\ee with
$\ppi$ and $\psi^\dagger$ the dual momenta as given in
(\ref{candu}). The space of all wave-functions must be projected
to a physical sub-space by imposing the constraints (\ref{const}).
The action of the operators in (\ref{const}) on wave functions
becomes more manifest after writing them as \be {\cal G} = [\,
\Phi,\Pi\, ] \, , \qquad \qquad \bar{\cal G} \, =
\, 
[\, \Phi,\bar{\Pi}\, ] \, . \ee We will impose the physical state
conditions in the weak form \be \label{four} {\cal G}\, |
\Psi_{\rm phys}\rangle \, =\, 0\, , \qquad \qquad \langle
\Psi_{\rm phys} | \, \bar{\cal G} \, = \, 0 \, . \ee It is not
possible to impose that both ${\cal G}$ and $\bar{\cal G}$
annihilate $|\Psi_{\rm phys}\rangle$, because they have a
non-trivial anti-commutation relation of the form $ \{{\cal
G}(\epsilon),\bar{\cal G}(\bar{\epsilon})\} \, =\, \Tr\Bigl(
[\epsilon,\Phi][\bar \epsilon,\Phi]\Bigr)$. 
States of the form $\bar{\cal
G}|\Psi\rangle$ and $\langle \Psi| {\cal G}$ are spurious: they
are orthogonal to physical states.

We would like to give an explicit description of all
physical states. It is useful to make the transition to a
component notation. The constraint operator ${\cal G}$ has an
expansion
 \ba & &{\cal G}\,  = \, F + \theta \, G + \bar\theta\,  K + \theta
\bar\theta \, L \nonumber \\[2.5mm]
{F}
 \is  [ \psi^\dagger,\phi ] 
, \ \ \ \qquad \qquad
\ G  =  
[\ppi,\phi] + [\psi^\dagger,\psi] 
\, , \\[2mm] 
J 
 \is  [ \psi^\dagger,\psi^\dagger ] 
\, , \ \ \ \qquad  \qquad 
L = [\, \psi^\dagger,\ppi 
] \nonumber 
\ea The
superspace equation (\ref{four}) thus implies four separate
physical state conditions, one for each generator.
These generators all commute
among each other, and form a supermultiplet: $\{
Q,{F}\} =\! G \! , \, $ $\{ Q^\dagger\! ,{F}\} \! =\! J \, $, and $ \{Q,J\}
= L\,$. The full set of
constraints thus defines a consistent supersymmetric truncation of
the matrix quantum mechanics.

Let $U$ be the bosonic unitary matrix that diagonalizes the
bosonic component $\phi$ of the matrix superfield. In general the
fermionic matrices $\psi$ and $\psi^\dagger$ will not be
diagonalized by $U$. Nonetheless, we can define \be (U\Phi
U^\dagger)_{kk} = z_k + \bar\theta\,\psi_k +
\psi_k^\dagger\,\theta +
 \bar\theta\theta f_k \, . 
\ee Here the $z_k$ are the $N$ eigenvalues of $\phi$. We will now
show that the full set of conditions (\ref{four})
is solved by physical states that depend on $z_k$ and
$\psi_k$ only. So in particular, they do not depend on the
off-diagonal fermionic fields: \be \label{new} \psi^\dagger_{kl}\,
|\Psi_{\rm phys}\rangle \, = \, 0 \qquad \qquad k \neq l\, . \ee

We first note that the bosonic gauge invariance condition $G=0$
implies that states are independent of $U$. Thus we may set $U=1$.
Next consider the constraint $F=0$. The off-diagonal components of
the fermionic matrix $\psi$ transform under $F(\eta)=\Tr(\eta F)$
as \be \label{ftrans} \delta_\eta \psi_{kl} = (z_k-z_l)\,
\eta_{kl}
\, .
\ee
Invariance under $F$ thus requires that states are independent of
$\psi_{kl}$ with $k\neq l$. This is the condition (\ref{new}).
Such states automatically satisfy the third constraint $J=0$.
To see that they are also annihilated by $L$, it is useful to
decompose $\pi_{kl}$ as \be
  \ppi_{kk} = 
     -{i{}}{\partial\over{\partial  z_k}} \, , \qquad \quad
    \ppi_{kl} 
     = 
       {-i {}}\,  {1\over (z_k -  z_l) }\; {\partial\over \partial U_{kl}}, \quad
       {{\rm for}\ k \neq l}\, .
\ee
The diagonal part $\ppi_{kk}$ does not
contribute to $L$, when acting on states that do not depend on
$\psi_{kl}$ with $j\neq k$. How about the off-diagonal part? Since
$\psi_k = (U\psi U^\dagger)_{kk}$, it does not commute with
$\ppi_{kl}$. Taking this into account, we find that on physical
states: \be \label{pij}
\ppi_{kl} 
\, = - i\; {\psi_{kl} \,(\psi^\dagger_k - \psi^\dagger_l)\over
(z_k-z_l)} \;
\, , \ \qquad {\rm for} \ k \neq l\, . \ee Now $L$ contains a term
$(\psi^\dagger_k-\psi^\dagger_l)\ppi_{kl}$, but as seen from
(\ref{pij}), this term also vanishes since it contains a factor
$(\psi^\dagger_k-\psi^\dagger_l)^2=0$. Hence $L=0$ on physical
states.

This result, that the subspace of singlet wave functions
$\Psi(z_k, \psi_k)$ satisfy a supermultiplet of gauge conditions
(\ref{four}), shows that it defines a consistent supersymmetric
truncation. One can also verify directly that it is invariant
under supersymmetry transformations: from (\ref{pij}) we deduce
that in the physical subspace, the supercharges reduce to \be
\label{one}
  Q = \sum_k \psi_k^\dagger 
  {\partial\over{\partial
 z_k}} \, , \qquad \qquad 
 Q^\dagger = \sum_k \psi_k\Bigl( {\partial\over{\partial
 z_k}} 
 + \sum_{l\ne k}{1\over{ z_k -  z_l}} \Bigr).
\ee This form for the supercharges on the singlet states was first
obtained in \cite{atish}\cite{arod}.

 By requiring that $Q^\dagger$
represents the hermitean conjugate of $Q$, we deduce the
innerproduct on the space of physical states involves a measure
factor \be \label{volb} {\rm Vol}\, {\cal C}_B(z) =
\prod_{i\neq j}(z_{i} - z_{j})\, = \Delta^2(z) . \ee with
$\Delta(z)$ the Vandermonde determinant. Here ${\cal C}_B(z)$
denotes the conjugacy class, under bosonic gauge transformations,
of the diagonal matrix $\phi$ with eigenvalues $z_k$. The
appearance of this measure factor is completely as expected. It is
convenient, however, to absorb a factor  of $\Delta$ into our wave
functions \be \label{split} \Psi(z,\psi) = \Delta(z)
\tilde{\Psi}(z,\psi)\, . \ee Note that since $\Delta(z)$ is
completely anti-symmetric under interchange of any pair of
eigenvalues $z_k$ and $z_l$, the new wave function has the
opposite statistics as the original one. (We will discuss the
statistics of the wave functions in a moment.) The new
supersymmetry generators now take the manifestly hermitian form
\be \label{hermit} Q = \sum_k \psi_k^\dagger\Bigl(
{\partial\over{\partial
 z_k}} +
       {{\partial W( z)}\over{\partial z_k}}\Bigr)\, , \qquad \quad
 Q^\dagger = \sum_k \psi_k\Bigl( {\partial\over{\partial
 z_k}}
             - \,{{\partial
             W( z)}\over{\partial z_k}}\Bigr),
\ee where \be \label{eff} W(z) = 
- \sum_{k<l} \log (z_k-z_l)\, . \ee
This corresponds exactly to the superpotential of the
supersymmetric Calogero-Moser model \cite{freedman}.

\medskip

\newsection{Super Virasoro symmetry of the Calogero model}

In this section, we give a short description of some properties of the
supersymmetric Calogero-Moser model \cite{cmtricks}.
The most interesting result
from our perspective is that, like its proposed dual string theory,
it exhibits a symmetry under a full ${\cal N}=2$ super-Virasoro algebra
\cite{bergvas}.

\medskip

\bigskip

\noindent
{\sc Spin and Statistics}

\def\downvac{|{\downarrow\downarrow\cdots\downarrow}\rangle}

\def\upvac{|{\uparrow\uparrow\cdots\uparrow}\rangle}

Since it is possible via $U(N)$-rotations to interchange any
eigenvalue superfield $(z_{i},\psi_{i})$ with any other eigenvalue
superfield $(z_{j},\psi_{j})$, the matrix wavefunctions should be
symmetric under this exchange operation. However, since our
wave-functions depend on anti-commuting variables, the model will
inevitably contain bosonic as well as fermionic sectors.

We can think of the system as $N$ particles moving in one
dimension, each with an internal spin $1\over 2$ degree of
freedom, a spin ``up" or ``down'' such that \be{ \psi^\dagger_{}
\downvac = 0, \ {\forall} \ i\ \ ; ~~~ \psi_{} \downvac \equiv |{
\underbrace{\downarrow\cdots\downarrow}_{i-1}
\uparrow\downarrow\cdots \downarrow}\rangle  ; ~~~ {\rm etc}...}
\ee Let us now define a bosonic and a fermionic interchange
operation with the property that \ba K_{ij} \, z_{i} = \, z_{j}\,
K_{ij}\, ,\quad & & \quad K_{ij} \, z_k = \, z_k \, K_{ij}\, ,
\qquad \qquad \quad  \nonumber \\[2mm]
  \kappa_{ij} \, \psi_{i}\, = \, \psi_{j} \, \kappa_{ij}\, , \
\quad & &\quad \kappa_{ij}\,  \psi_k =\, \psi_k \, \kappa_{ij}\,
,\ \qquad {\mbox{\small $k\neq i,j$}}. \ea An explicit
representation of the fermionic interchange operator is \cite{cmtricks}
\be
\label{kij} \kappa_{ij}\equiv
1-(\psi_{i}-\psi_{j})(\psi^{\dagger}_{i}-\psi^\dagger_{j})\, . \ee
In addition to interchanging the $i$ and $j$ spin state,
$\kappa_{ij}$ also multiplies the overall wavefunction by a minus
sign in case both spins point in the up-direction. This minus sign reflects
the Fermi statistics of $\psi_{i}$ and $\psi_{j}$.

We can now define the total exchange operation as the product of
the bosonic and fermionic one \be \mbox{\Large $\kappa$}_{ij} =
K_{ij}\, \kappa_{ij}. \ee Acting with $\mbox{\Large
$\kappa$}_{ij}$ amounts to interchanging particle $i$ with
particle $j$. We can thus specify the overall statistics of the
physical wavefunctions by means of the requirement that \be
\label{statistics} \mbox{\Large $\kappa$}_{ij}\,
|\tilde{\Psi}_{\rm phys}\rangle = -|\tilde{\Psi}_{\rm phys}\rangle
\, \qquad \qquad \forall \ {i,j} \ee The minus sign on the
right-hand side ensures that the original wavefunction, before
splitting off the Vandermonde determinant (see eqn (\ref{split})),
is completely symmetric. The condition (\ref{statistics}) implies
that particles with spin ``up" are fermions, while particles with
spin ``down" are bosons. We can call this the spin-statistics
theorem for our model.

\medskip

\bigskip

\noindent {\sc Diagonalization of the Hamiltonian}

We now summarize the algebraic method by which one can obtain the
energy spectrum of the supersymmetric Calogero-Moser model. For a
more detailed discussion, we refer to the original papers
\cite{freedman}\cite{poly}\cite{cmtricks}\cite{ghosh}.

Instead of the original Hamiltonian (\ref{ham}), we will choose to
consider
\be \label{calh}\CH = \PP + \omega_0 \JJ + \omega_0^2 \, \KK\, .
\ee Here we are making use of the natural freedom in
super-conformal quantum mechanics to choose different
time-slicings \cite{DFF}\cite{andyl}. As we will see, the above
Hamiltonian ${\cal H}$ will be most naturally identified with the
global time coordinate of $AdS_2$.

Using the explicit form of the supercharges (\ref{hermit}), a
straightforward calculation gives that \be \CH = {1\over 2}
\sum_{i=1}^N \left(\, p_i^2+ \omega_0^2 z_i^2 + \omega_0
\psi^\dagger \psi \right) +  \sum_{i<j} {1- \kappa_{ij} \over
(z_{i}-z_j)^2}, \ee with $\kappa_{ij}$ the fermionic exchange
operator defined in (\ref{kij}). We see that the Hamiltonian
describes a system of interacting eigenvalues. In effect, the
interaction is such that eigenvalues always repel each other: it
represents a $2/r^2$
repulsion between the boson states with $\kappa_{ij}=-1$, and
although it vanishes between two fermionic states with
$\kappa_{ij}=1$, such particles still avoid each others presence
since their wavefunctions are anti-symmetric.

In case all the
eigenvalues have spin down, so that all $\kappa_{ij}=1$, the
Hamiltonian reduces to a decoupled set of one-particle
Hamiltonians. The ground state wave function in this case reads: \be
\label{fnot} f_0(z) =
\Delta(z) e^{-{1\over 2} \omega_0\sum_{i=1}^N z^2_i} \, . \ee This
state represents the filled Fermi sea of the first $N$ energy
levels.

A very useful set of operators are the so-called Dunkl operators
(here $z_{ij} = z_i-z_j$) \be
\label{dunkl} D_j= {\partial\over
\partial z_j}+  \sum_{k\neq j}{1\over z_{jk}}(1-\kappa_{jk})\, .
\ee These operators are instrumental in demonstrating the
integrability of the Calogero-Moser model, and in the explicit
construction of its energy spectrum. They can be shown to satisfy
the useful property that they all commute with each other
$[D_i,D_j]=0\,.$ 

Now following \cite{cmtricks}, we transform the Hamiltonian via
\be \label{traf} \bar{{\CH}}\equiv
    f_0(z)^{-1}\, {\CH}\, f_0(z)
\, . \ee  A straightforward calculation shows that the transformed
Hamiltonian can be written in the following compact form
\footnote{Here we drop a constant term $E_0 = {1\over 2}
N\omega_0$.}
\be \label{newham}
\bar{{\CH}}\,
=\, \omega_0 L_0 
- \Delta_2 
\vspace{-5mm} \ee where \be \label{lnot} L_0 = \sum (\,z_i
\partial_i \, + \, {1\over 2} \psi_i^\dagger \psi_i) , 
\, , \qquad \qquad \Delta_2 = {1\over 2} \sum D_i^2 \, . \ee
The new Hamiltonian (\ref{newham}) still looks like a complicated
operator. However, we can now use that the Dunkl operators $D_j$
are homogeneous of degree $-1$ under overall rescalings of the
eigenvalues, so that \ba \label{scale} [L_0 , D_j] = - D_j\, ,
\qquad \qquad [L_0,\Delta_2]= -2 \Delta_2\, . \ea This observation
essentially trivializes the computation of the spectrum, since it
allows us to rewrite $\bar \CH$ as follows \cite{cmtricks}
\ba\label{smart2} \bar \CH \, = \, e^{-{1 \over
2\omega_0}\Delta_2}\; \omega_0 L_0 \, e^{{1\over
2\omega_0}\Delta_2}\, .\ea Hence, up to conjugation, the
Hamiltonian looks like the sum of harmonic oscillator Hamiltonians
for the individual eigenvalues.

\medskip

\bigskip

\noindent
{\sc Super Virasoro Symmetry}

As already suggested by our notation, it turns out that the
operator $L_0$ given in (\ref{lnot}) is one of an infinite set of
conserved charges that generate a Virasoro algebra. Furthermore,
by using the $Osp(2|2)$ supersymmetry, one can exhibit a full
${\cal N}\!=\! 2$ super Virasoro algebra \cite{bergvas}. The
construction of these charges involves a
supersymmetric generalization of the Dunkl operator, given by
\be {\cal D}_i = {\cal D}_i^0 + \sum_{i\neq j} {\psi_{ij}\over
z_{ij}}\, (1-\kappa_{ij}) \, , \qquad \quad {\cal D}_i^0 =
{\partial\over
\partial\psi_i} + \psi_i {\partial\over \partial z_i} \, . \ee
Here $\psi_{ij} = \psi_i-\psi_j$. 
Using this supersymmetric differential operator,
the ${\cal N}=2$ generators are obtained via
\cite{bergvas}
\ba L_\xi \is \sum_i ({\cal D}_i \xi \, {\cal D}_i - {1\over
2}({\cal D}_i^0 \xi_i){\cal D}_i
+\lambda \xi'_i)\, , \nonumber \\[1.5mm]
 G_\epsilon \is \sum_i (\epsilon_i {\cal D}_i + \lambda ({\cal D}^0_i\epsilon_i))
\ea
where $\xi(z,\psi)$ and $\epsilon(z,\psi)$ are arbitrary superfields,
and $\lambda$ denotes the conformal dimension of the ground state.
If the Calogero interaction is turned off, these charges reduce
to the standard form of the ${\cal N}=2$ generators expressed
in ${\cal N}=1$ superspace. The $L_0$ operator is obtained by taking
$\xi_0= z$. This gives $L_{\xi_0}= \psi L_0$, with $L_0$ as in
(\ref{lnot}). Hence if we conjugate the above generators as in (\ref{smart2}),
we obtain a super-Virasoro symmetry in which (\ref{calh}) acts like
the $L_0$-generator.

\newsection{Calogero interaction from GS action}

Let us recapitulate. Based on the classical form of the
Green-Schwarz action, we have argued that physical states of IIA strings
on $AdS_2$ are primary operators of a target space
${\cal N}\! =\! 2$ Virasoro algebra. The same symmetry naturally appears
in the Calogero-Moser model that describes the eigenvalue dynamics of the
superconformal matrix model. If we want to interpret the matrix model
as the holographic dual CFT${}_1$, correlation functions of primary
operators in the CFT${}_1$ should correspond to scattering amplitudes
of single IIA string states at the $AdS_2$ boundary.
Such states correspond to local vertex operators on the world-sheet.
Schematically, we are looking for a correspondence \cite{jan}
\be
\label{ident}
\Bigl\langle\, \prod_i \int \! d^2 z_i \; V_i(z_i,\XX_i) \, \Bigr\rangle_{\rm worldsheet} =
\Bigl\langle
\, \prod_i V_i(\XX_i)\,
\Bigr\rangle_{\rm CFT_1}.
\ee
Here the vertex operator $V(z,\XX)$ is assumed to contain a factor
$\delta(z-\XX)$ that attaches the string world-sheet at the point $z$
to the boundary point $\XX$. The expectation value on the left-hand
side involves a sum over all connected and disconnected worldsheets
that interpolate between the points $\XX_i$.

Now, while the appearance of the ${\cal N}\!=\! 2$ superconformal symmetry on
both sides is encouraging, it would clearly be desirable to find a somewhat
more decisive correspondence, such as a derivation of the Calogero interaction
from the IIA string theory. In this section we will point to a promising route
towards such a derivation.

One of the most central objects in the description of the Calogero dynamics
is the Dunkl operator (\ref{dunkl}). It encodes, as well as decodes, the
complete Calogero interaction. We will now show how the exact same operator
naturally appears in the IIA worldsheet theory.

Let us return to the GS action in the form (\ref{newsig}). We will again
ignore the interaction term. The remaining action reduces to the free
field representation of a $gl(1|1)$ current algebra with generators
\cite{guruswamy}
\ba
g_- = \ppi \TTheta\, , \qquad g_+ = \TTheta^+ \XX \, , \qquad
h_1 = \TTheta^+ \TTheta\, , \qquad
h_2 = \ppi \XX \, .
\ea
The stress-tensor is given by the corresponding Sugawara
expression \be T = {1\over 2}(h_1^2 - h_2^2 - g_-g_+ + g_+g_-)
\equiv c_{ab}j^{\, a} j^{\, b} \ee Here have done a $U(1)$ twist
such that all fields have
spin $1/2$, so that the 
currents are non-anomalous. This relation between the current
algebra and the stress-energy tensor results in a differential
equation for (unintegrated) correlation functions of current
primary fields, given by the Knizhnik-Zamolodchikov equation: \be
\label{kz} \Bigl({\partial \over \partial z_i} + \sum_{j\neq i}\,
{c_{ab}t^a_i t^b_j \over z_i-z_j}\, \Bigr) \, \Bigl\langle\,
\prod_i V_i(z_i) \, \Bigr\rangle \; =\;  0
\ee
Here $t_i^a$ is the $gl(1|1)$ Lie-algebra generator associated with the
current $j^{\, a}$, acting on the vertex operator $V_i(z_i)$.

The KZ differential operator already looks very similar to the
Dunkl operator (\ref{dunkl}). In fact, they exactly coincide: the
$gl(1|1)$ Lie-algebra (here we are using the same letter for the
zero-mode as for the corresponding current) \be
\anti{g_-}{g_+}=h_1-h_2 \, , \qquad [h_{1},g_\pm] =[h_2,g_\pm]=
\pm g_\pm \, \qquad g_\pm^2 = 0\, , \ee has a simple
representation in terms of a fermionic creation and annihilation
operator $\{\psi^\dagger,\psi\}=1$ via the identification
\be g_+ \leftrightarrow \psi^\dagger, \qquad g_- \leftrightarrow
\psi \, , \qquad h_1 - h_2 \leftrightarrow 1 \, \qquad h_1 + h_2
\leftrightarrow 2 \psi \psi^\dagger  \, . \ee With this dictionary
\cite{ahn2} \be c_{ab} \, t_{\, i}^{\; a} t_{\, j}^{\; b} =
(\psi_i-\psi_j)(\psi^\dagger_i - \psi^\dagger_j) = 1-\kappa_{ij}
\ee with $\kappa_{ij}$ the fermionic interchange operator
(\ref{kij}). This shows that the operators in (\ref{kz}) and
(\ref{dunkl}) are the same. This exact correspondence does not yet
prove the holographic identification (\ref{ident}), since the
dictionary is still incomplete and interactions should be
included. (The Yangian structure discussed in \cite{ahn2} may be
of help here.) Still, it provides yet another encouraging sign
that we are on the right track.

\newsection{Open Issues}

We have presented several pieces of evidence that
IIA superstring theory on $AdS_2$ admits a holographic dual in
terms of superconformal matrix quantum mechanics. The most attractive
feature of this duality is that both systems possess a large symmetry
structure, which should allow even more precise checks and possibly even
a derivation. In this sense, it forms a useful playground for testing the
recent ideas on integrability in 4-d SYM and its string dual.
There are still a many unresolved questions, however.

\bigskip

\noindent
{\it Interactions}

A first technical challenge is to include interactions, both in the
sigma model calculations as well as in the matrix model. In particular,
one needs a precise worldsheet derivation of the critical $AdS_2$
radius for which the string theory is consistent. Further, in the
dictionary with the matrix model, it is of importance to
find the proper identification of the string coupling. Our proposal,
based on lessons from the type 0A matrix model \cite{hat}, is that
it is related to the logarithmic superpotential $W_0=q\log \Phi$
via $g_s = q^{-1}$. Since the matrix degrees of freedom are to be
thought of as D-particles in non-critical IIA string theory, there are
undoubtedly useful lessons to be learned from studying their spectrum
and interactions.

\bigskip

\noindent
{\it $AdS_2$ fragmentation}

Assuming that the Calogero-Moser model captures the proper
low energy dynamics of the D-particles, it provides a quite
concrete realization of the idea of $AdS$-fragmentation \cite{fragm}.
A localized cluster of many eigenvalues, from the perspective of a
distant probe eigenvalue, represents a $q/z^2$ repulsive potential.
So we can think of the cluster as creating an $AdS_2$ space with
some RR-flux $q$. When the probe approaches the cluster, it can
resolve the different constituents. In terms of space-time physics,
this looks as if the $AdS_2$ fragments into many $AdS_2$ space-times,
each with a fraction of the total RR flux.

\bigskip

\noindent
{\it Non-critical M(atrix) theory}

A natural question is whether,
like in the critical IIA theory, the D-particles have a dual
interpretation as particles with non-zero Kaluza-Klein momentum
in a third, M-theoretic direction.
The $AdS_2$-background (\ref{adst}), when assembled into a
3-geometry via the standard KK Ansatz, represents a DLCQ-like
reduction  of $AdS_3$ along a light-like $y$-direction, \cite{andy}.
Non-critical M-theory on $AdS_3$, if it exists, probably has
a holographic dual given by a 2-D superconformal field theory;
it is tempting to identify this SCFT with the continuum limit
of the Calogero model at $W_0=0$.

\bigskip
\bigskip

\begin{center}
{\bf Acknowledgements}
\end{center}
\medskip
I would like to thank J. McGreevy and S. Murthy for collaboration
on the material in sections 4 and 5. I further benefited from
helpful discussions with O.~DeWolfe, L.~Dolan, N.~Itzhaki,
I.~Klebanov, J~.Maldacena, D.~Malyshev, A.~Polyakov, L.~Rastelli,
N.~Seiberg, A.~Sovolyov and E.~Witten. This material is based upon
work supported by the National Science Foundation under grants
No.~0243680. Any opinions, findings, and conclusions or
recommendations expressed in  this material are those of the
authors and do not necessarily reflect the views of the National
Science Foundation.

\renewcommand{\Large}{\large}

\end{document}